\documentclass[12pt,a4paper]{article}
\usepackage{amsmath,amsfonts,latexsym,amssymb,amsthm}
\usepackage{curves}

\begin{document}
  
%For lines of pictures:
\setlength{\unitlength}{0.4pt}
\linethickness{0.15mm}          

\theoremstyle{plain}
\newtheorem{thm}{Theorem}[section]
\newtheorem{lem}{Lemma}[section]

\theoremstyle{definition}
\newtheorem{defn}{Definition}[section]
\newtheorem{case}{Case}

\renewcommand{\thesection}{\Roman{section}}
\renewcommand{\H}{\mathcal{H}}
\newcommand{\Tr}{\operatorname{Tr}} 
\newcommand{\defeq}{\overset{ \text{def} }{=}}
\newcommand{\hodge}{*\!\!\;}
\newcommand{\mint}{-\!\!\!\!\!\!\int}
\newcommand{\mslash}{/\!\!\!}
\newcommand{\dirac}{\gamma^\mu\nabla_\mu}
\newcommand{\SO}{{\rm SO}}
\newcommand{\Spin}{{\rm Spin}}
\newcommand{\Cliff}{{\rm Cliff}}
\newcommand{\WF}{{\rm WF}}
\newcommand{\bRe}{{\rm Re}}
\newcommand{\bIm}{{\rm Im}}
\newcommand{\supp}{{\rm supp}}
\newcommand{\spec}{{\rm spec\,}}
\newcommand{\car}{{\rm CAR}}
\newcommand{\Pol}{{\rm WF}_{pol}}
\newcommand{\bS}{{\mathcal S}}
\newcommand{\bF}{{\mathcal F}}
\newcommand{\bM}{{\mathcal M}}
\newcommand{\bA}{{\mathcal A}}
\newcommand{\bK}{{\mathcal K}}
\newcommand{\bH}{{\mathcal H}}
\newcommand{\bO}{{\mathcal O}}
\newcommand{\bI}{{\mathcal I}}
\newcommand{\bD}{{\mathcal D}}
\newcommand{\bN}{{\mathcal N}}
\newcommand{\bQ}{{\mathcal Q}}
\newcommand{\bL}{{\mathcal L}}
\newcommand{\bE}{{\bf E}}
\newcommand{\bR}{{\bf R}}
\newcommand{\bZ}{{\bf Z}}
\newcommand{\bC}{{\bf C}}
\newcommand{\ri}{{\rm i}}
\newcommand{\rd}{{\rm d}}
\hyphenation{Gra-du-ier-ten-kol-leg Ele-men-tar-teil-chen}

\title{Adiabatic Hadamard States for Dirac Quantum Fields on Curved Space}
\author{Stefan Hollands\footnote{Electronic mail:
        hollands@physik.hu-berlin.de}\hspace{0.5em}\\
       \it{Department of Mathematics, University of York,} \\ 
       \it{York YO10~5DD, UK}}
\date{}

\maketitle

\begin{abstract} 
In this paper we propose a definition of quasifree Hadamard states for
spinor fields on a curved space-time by specifying the 
Polarisation Set of the two-point function. 
We prove that the thermal equilibrium state on an ultrastatic 
space-time is Hadamard.
We then construct an adiabatic vacuum state on a general globally hyperbolic
Lorentz manifold using a factorisation of the spinorial Klein-Gordon operator. 
This state is pure. In what constitutes the main part of the paper, 
we show that it is also Hadamard. As a side result, we obtain the 
propagation of singularities of the spinorial Klein-Gordon operator.
Some notation and results are collected in the Appendix.
\end{abstract}

\pagebreak

\section{Introduction}

In many cases of physical interest, for example the early stages
of the universe or stellar collapse, one faces the problem of
constructing quantum field theories on a non-static curved space-time.
As a preparation for more complicated models such as QED, we shall 
study adiabatic quantum states for a free Dirac field on a 
general globally hyperbolic space-time. 

We find it convenient to work in the algebraic 
framework of quantum field theory, which started with the work
of R. Haag and D. Kastler \cite{haka}, for an overview see \cite{haag}. 
In this approach one deals with a net of $C^*$-algebras
$\{\bA(\bO)\}_{\bO \subset M}$ of observables localised in a 
space-time region $\bO \subset M$. The algebra $\bA = 
\overline{\cup_{\bO \subset M} \bA(\bO)}$ is called the `quasilocal
algebra'. In this approach, quantum states are positive
normalised linear functionals on $\bA$. One of the major difficulties
of QFT on curved space-times is to pick out physically reasonable
states. This is because, due to the absence of space-time symmetries, 
there is no analogue of the spectrum condition, which is a powerful 
tool to single out physical states in Minkowski space.  

Hadamard states are thought to be good candidates for physical states at
least for free quantum field theories in curved space-time. They allow
for a point-splitting renormalisation of the stress-energy tensor
$T_{\mu\nu}$ \cite{Wald}. R. Verch \cite{Verch} has
shown that in case of the quantised Klein-Gordon field, Hadamard states are
quasi-equivalent and he has also shown local definiteness in the 
sense of Haag et. al. \cite{HNS}. 

Numerous papers have been devoted to the study of Hadamard states for 
free scalar field theories, especially since the important discovery, 
of M. Radzikowski \cite{Rad1}, that (quasifree) Hadamard
states can be characterised by the Wave Front Set of their 
two-point function. This made it possible, among other things,
to construct Wick products of free fields 
or to adapt the Epstein-Glaser approach to renormalisation to 
theories on a curved background, using the powerful
tools microlocal analysis and the theory of pseudo-differential 
operators (PDO's) \cite{fred}. 

It seems that less work has been done for spinor fields
in this direction. This is not due to conceptual problems 
but rather because the microlocal analysis of 
multicomponent fields is technically more involved. The extension
of the techniques developed for scalar fields to multicomponent
fields seems desirable. We propose to characterise Hadamard states 
in terms of their Polarisation Sets, a concept which is a refinement of
the Wave Front Set of a distribution \cite{Denck}.

If the space-time in question has a timelike Killing vector field, one
can fix a ground state by projecting on the positive frequent solutions
of the Dirac equation. This strategy is however not appropriate on a 
general globally hyperbolic space-time, because positive and negative
frequent modes (determined at an instant of time) 
will mix when propagated. Or, to put it differently, the Hamiltonian is 
not diagonal with respect to positive and negative frequent modes. 
Instead, they have to be determined dynamically off the Cauchy surface. 
This is achieved by a factorisation of the spinorial Klein-Gordon 
operator into positive and negative frequency parts, which has been
considered before in W.~Junker's work on adiabatic vacuum states for the
Klein-Gordon field \cite{Junk}. Making use of a characterisation 
of quasifree states on the $\car$-algebra \cite{araki}, we obtain a
pure state which approximates the vacuum state of an ultrastatic 
space-time. For this reasons, we will call it an adiabatic ground state.
The main result (Thm. \ref{mainthm}) of our work is that it is 
of Hadamard type. As a warm-up 
exercise, we first obtain ground and thermal equilibrium states 
on ultrastatic spacetimes, and show that these are Hadamard. The analysis
relies heavily on various results from microlocal analysis and the 
theory of PDO's, especially the propagation 
of singularities theorem. As side-result, the propagation of 
singularities for the spinorial Klein-Gordon operator is obtained in 
the proof of Thm. \ref{mainthm}.
Our techniques may be used to prove that the states 
introduced by L. Parker \cite{parker} for a free Dirac field on a 
Robertson-Walker space-time are of Hadamard type. This was already conjectured 
in \cite{wellmann} and will be discussed in more detail in a 
subsequent paper. Recently, adiabatic states describing thermal
equilibrium have been invented for the free Klein-Gordon field 
on a Robertson-Walker space-time \cite{trucks}. It would be 
interesting to ask whether this can be done for the Dirac field as well. 

Our work is organised as follows. In Sec. \ref{def} we define the 
local algebras of observables corresponding to a free Dirac field
and characterise quasifree states. In Sec. \ref{haddef} we give 
present the Hadamard condition for the Dirac field. In the next
three sections, we give a definition of ground, KMS and 
adiabatic states and prove that the Hadamard condition holds in 
all three cases. We have shifted some results and definitions 
from microlocal analysis into the Appendix, where further reference 
can be found. 

It should be noted that independent of our work, M. Radzikowski
has investigated a similar definition of Hadamard states for
spinor fields and also considered the propagation of singularities 
\cite{Rad2}. We are very grateful to him for making his 
results avaliable to us prior to publication.

\section{Local algebras and quasifree states}\label{def}

It was shown in \cite{Dim1} how to associate a net of algebras
of observables to the free Dirac field on a globally hyperbolic
spacetime. For convenience of the reader we shall
briefly sketch the main line of argument and recall how a 
certain class of states on this algebra, the so-called quasifree
states, can be characterized. This will be used in order to construct
quasifree Hadamard states in the later sections.  

Let us start with some notation. Throughout this article $(M,g)$ will
denote a $4$-dimensional globally hyperbolic Lorentz manifold of
signature $(+,-,-,-)$. Globally hyperbolic manifolds can be foliated by 
$3$-dimensional spacelike hypersurfaces $\Sigma$, implying that topologically
$M = \Sigma \times \bR$. It can be shown that every globally 
hyperbolic Lorentz manifold admits a spin-structure, i.e. a $2:1$ 
cover of the frame-bundle with a $Spin(3,1)$ principal fibre-bundle
$\cal P$ \cite{Geroch}. Spinors resp. cospinors are sections in the associated
vector bundles  
\begin{eqnarray*}
DM = {\cal P} \times_\tau E, \quad D^*M = {\cal P} \times_{\tau^*} E^*, 
\end{eqnarray*}
where $\tau$ denotes the fundamental
representation of $Spin(3,1)$ on the representation space $E$
and $\tau^*$ is the conjugate representation on the dual $E^*$. 
Gamma matrices on a spin-manifold are
defined to satisfy the usual anticommutation relations\footnote{
Indices from the Greek alphabet denote components
in a local chart, while indices from the beginning of the
Roman alphabet label components in a local frame. Letters from the 
middle of the Roman alphabet mean spacelike components with respect to
a local chart of $\Sigma$.}
$\gamma^\mu\gamma^\nu + \gamma^\nu\gamma^\mu = 2g^{\mu\nu}$
and contraction of vector indices with gamma matrices is denoted by a slash.
The pull-back of the Levi-Civita-connection to the spin-cover of 
$(M, g)$ is called the spin-connection and will be denoted by $\nabla$. 
The connection coefficients can be expressed in terms of the 
Christoffel symbols of the Levi-Civita connection. In a local frame
we have
\begin{eqnarray*}
\nabla_\mu = \partial_\mu -\frac{1}{4} \gamma^a\gamma^b \Gamma_{ab\mu}. 
\end{eqnarray*}
We consider the Dirac equation for smooth spinor and cospinor fields
$u$ resp. $v$, 
\begin{eqnarray}\label{diraceq}
(i\dirac - m)u = 0, \quad v(-i\dirac - m) = 0
\end{eqnarray}
and construct the algebra $\car(\bK, \Gamma)$ associated to the 
space of classical solutions to this equation. Let $L^2(\Sigma, DM)$ and
$L^2(\Sigma, D^*M)$ be the spaces of square integrable spinor resp.
cospinor fields equipped with the scalar product
\begin{eqnarray*}
\langle u_1, u_2 \rangle = \int_\Sigma (\bar u_1\mslash n
u_2)(x)|h(x)|^{1/2}d^3 x, 
\end{eqnarray*}
where $n$ is the forward directed normal vector field to $\Sigma$, bar 
denotes the Dirac conjugate and $h$ is the induced metric on $\Sigma$. 
Let $\bK = L^2(\Sigma, DM) \oplus L^2(\Sigma, D^*M)$ and define 
the antiunitary involution ($\Gamma^2 = 1$) on $\bK$, 
\begin{eqnarray*}
\Gamma = 
\left[
\begin{matrix}
0 & \iota^{-1}\\
\iota & 0
\end{matrix}
\right], 
\end{eqnarray*}
where $\iota: DM \to D^*M$ is the antilinear map of Dirac conjugation.
$\bK$ is the space of initial data to the Dirac equation for particle
and antiparticle fields.
The algebra $\car(\bK, \Gamma)$ is the unique $C^*$-algebra generated
by elements $A(F), \,\, F \in \bK$ obeying the anticommutation relations
\begin{eqnarray*}
\{A(F_1),A(F_2)\} = \langle 
\Gamma F_1, F_2 \rangle_\bK, 
\end{eqnarray*}
and with the property $A(F)^* = A(\Gamma F)$, see e.g. \cite{araki}. 

If $U$ is an isometry on $\bK$ commuting with $\Gamma$, then there
exists a $*$~-automorphism $\alpha_U$ of the $\car$-algebra satisfying
\begin{eqnarray*}
\alpha_U(A(F)) = A(UF).
\end{eqnarray*}
$\alpha_U$ is called the Bogoliubov automorphism corresponding to $U$.
In order to find the observables located in a space-time region one 
makes use of the one-to-one correspondence between initial data and
solutions to the Dirac equation. More formally, $\bK$ is isomorphic
to the completion of the space of solutions 
$C^\infty_0(M, DM)/ker S \oplus C^\infty_0(M,
D^*M)/ker S^*$, where $S$ resp. $S^*$ are the uniquely defined causal 
propagators for Eqs.~\eqref{diraceq}, see \cite{Dim1} (from now on we drop the
star on $S$). The isomorphism is given by 
restricting a solution the Cauchy surface, $Su \oplus Sv \to \rho Su
\oplus \rho Sv$, $\rho$ denoting the restriction operator to the surface
$\Sigma$. The isomorphism shows in particular that the construction is
independent of a particular choice for $\Sigma$. The field operators for
particle/antiparticles are given by
\begin{eqnarray*}
\Psi(v) = A
\left(
\left[
\begin{matrix}
0\\
i \rho S v
\end{matrix}
\right]
\right),
\quad
\bar \Psi(u) = A
\left(
\left[
\begin{matrix}
i \rho S u\\
0
\end{matrix}
\right]
\right),
\end{eqnarray*}
with the usual anticommutator
\begin{eqnarray}\label{car}
\{\Psi(v), \bar\Psi(u)\} = i S(v,u).
\end{eqnarray}
The local algebras of observables are
\begin{eqnarray*}
\bA(\bO) = C^*\{A(F), \quad F \in \bK, 
\quad supp(F) \subset \bO\}^{\rm even},  
\end{eqnarray*}
where we mean the even part of the $C^*$-algebra generated by these
elements, i.e. the fixalgebra of the even/odd 
automorphism generated by $F \to -F$. 

A state on $\bA$ is called 
{\it gauge invariant}, if 
$\omega = \omega \circ \alpha_\theta$, where 
$\alpha_\theta$ is generated by 
\begin{eqnarray*}
F = 
\left[
\begin{matrix}
f_1\\
f_2
\end{matrix}
\right]
\to 
\left[
\begin{matrix}
e^{i\theta}f_1\\
e^{-i\theta}f_2
\end{matrix}
\right]
\end{eqnarray*}
Moreover, a state $\omega$ such that  
\begin{eqnarray*}
\omega(A(F_1)\dots A(F_n)A(G_1) \dots A(G_n)) = det\left( \langle
\Gamma F_i, T G_j \rangle_\bK \right)_{i,j = 1,\dots,n}
\end{eqnarray*}
for some operator $0 \le T \le 1, \,\, T + \Gamma T \Gamma = 1$ is 
called {\it quasifree}. These states give rise to 
a Fock-representation $\pi$ of the $\car$ via a the GNS-construction,  
\begin{eqnarray}\label{representation} 
\pi(A(F)) = a(T^{1/2}F) + a(\Gamma (1-T)^{1/2}F)^*, 
\end{eqnarray}
where the $a$'s denote the creation and annihilation operators on the 
fermionic Fock-space $\bF$ over $\bK$. Conversely, any operator $T$ with
the above properties determines a quasifree state on $\bA$ via 
Eq.~\eqref{representation}. Such a state $\omega$ is pure if and only if $T$ is
a projection. Moreover, a quasifree state is gauge
invariant if and only if $T = B \oplus \iota(1-B)\iota^{-1}$ for some
operator $0 \le B \le 1$.

\section{The Hadamard condition for spinor fields}
\label{haddef}

Hadamard states were introduced in order to define the
expectation value of the stress energy tensor of a linear scalar 
quantum field on a curved space time. Their two-point functions 
have a singularity of a particular form on the diagonal, according
to the dimensionality of spacetime. For a mathematically 
rigorous definition see Kay and Wald \cite{kaywald}. 
In this article we prefer to work with a definition that emphasizes the 
microlocal properties of Hadamard states. Both characterisations are known to 
be equivalent since the work of M. Radzikowski \cite{Rad1}, at least for
scalar fields. Due to the vector character of Dirac fields, the definition is however 
technically more involved than for scalar fields. 
For a brief explanation of the technical ingredients and a few important
results we refer to the Appendix. 

Let $u \in C^\infty_0(M,DM)$ and $v \in C^\infty_0(M,D^*M)$. We denote by
\begin{eqnarray}\label{twopointfunctions}
\Lambda^{(+)}(v, u) = \omega(\Psi(v) \bar\Psi(u)), \quad  
\Lambda^{(-)}(v, u) = \omega(\bar\Psi(u)\Psi(v)) 
\end{eqnarray}
the spatio-temporal two-point functions of a state $\omega$, which we 
assume to be distributions. In order to lighten the notation, let us introduce the set
$C \subset T^*(M \times M)$ of all pairs 
$(x_1, \xi_1) \sim (x_2, \xi_2)$, where $\sim$ 
means that $x_1$ and $x_2$ can be joined by a null-geodesic $c$ 
such that $\xi_1 = \dot c(0)$ and $\xi_2 = \dot c(1)$.   
For later use, we also introduce the sets $N_\pm \subset T^*M$ 
of all future/past directed covectors $\xi$, i.e. satisfying
$g(\xi,\xi)\ge 0$ and $\pm\xi^0 \ge 0$.
\begin{defn}\label{Hadamard}
A quasifree state $\omega$ is said to be `Hadamard' if the (primed) 
Polarisation sets 
$\Pol'(\Lambda^{(\pm)})$ of its spatio-temporal two-point functions are of
the form 
\begin{eqnarray} \label{77}
\big\{ (x_1, \xi_1, x_2, \xi_2, w): 
(x_1, \xi_1) \sim  (x_2, \xi_2), \,\, \pm\xi_1^0 \ge 0, \,\,
w = \lambda\mslash\xi_1 \bI_\nabla(x_1, x_2) \big\},
\end{eqnarray}
where
\begin{eqnarray*}
w \in D_{x_1}M \otimes D^*_{x_2}M, \quad  \xi_1 \in T^*_{x_1}M, 
\quad \xi_2 \in T^*_{x_2}M, \quad \lambda \in \bR 
\end{eqnarray*}
and $\bI_\nabla$ is the parallel transport function on the spin-bundle. 
\end{defn}
To prove that a given quasifree state is of Hadamard type we only have
to investigate the Polarisation Set of its two-point function.
This shall be done for ground states and KMS states on ultrastatic 
space-times and adiabatic vacuum states on general globally hyperbolic 
space-times. Note that the above definition especially implies the
relation 
\begin{eqnarray*}
\WF'(\Lambda^{(\pm)}) = C \cap (N_\pm \times N_\pm)
\end{eqnarray*}
for the Wave Front Set.

\section{Ground states}\label{ground}

As we mentioned before, ground states can be defined properly on 
ultrastatic spacetimes only. For the free Dirac field, such 
a state can be constructed quite easily. An ultrastatic spacetime
is by definition a spacetime $M = \bR \times \Sigma$ with a line element
$d s^2 = d t^2 - h_{ij}d x^id x^j$, the spatial part of the metric 
being time independent. Eqs.~\eqref{diraceq} can be written as 
\begin{eqnarray*}
i\partial_t F = 
\left[
\begin{matrix}
H & 0\\
0 & -\iota H \iota^{-1}
\end{matrix}
\right]F=\widetilde{H} F
\end{eqnarray*}
on $\bK$, where 
\begin{eqnarray*}
H = -i\gamma^0\gamma^j \nabla_j + \gamma^0 m. 
\end{eqnarray*}
$H$ is an essentially self-adjoint operator on $L^2(\Sigma, DM)$. Let us
denote by $E_\pm$ the projections on the positive/negative spectral subspace
of $H$. A projector $T$ with the property
$T + \Gamma T \Gamma = 1$ is the given by 
\begin{eqnarray*}
T = 
\left[
\begin{matrix}
E_+ & 0\\
0 & \iota E_- \iota^{-1}
\end{matrix}
\right].
\end{eqnarray*}
By the remarks of the preceding section, it defines a pure state
on $\bA$, which is also gauge invariant. The definition of $T$ takes 
into account that particle 
states move forward in time with positive energy whereas antiparticle 
states move backwards in time with positive energy or move forwards 
in time with negative energy. It can be seen that the above state is 
the Fock-state which has the lowest energy among all Fock-states 
where the energy operator can be defined and has positive energy. 
The state defined by $T$ is therefore the uniquely defined
ground state. The following theorem is a special case of Thm. 
\ref{mainthm}.
\begin{thm}
The ground state on an ultrastatic space-time $(M,g)$ is Hadamard in the
sense of Def. \ref{Hadamard}.
\end{thm}

\section{Thermal equilibrium states}\label{KMS}

On ultrastatic space-times, thermal equilibrium states can be defined. 
Such a state is a KMS state \cite{HHW} whose modular automorphism
group coincides with the dynamics of the system. We define
\begin{eqnarray*}
T_\beta = \frac{\exp (-\beta \widetilde{H}/2)}{2\cosh (\beta
\widetilde{H}/2)} = \frac{\exp (-\beta H/2)}{2\cosh (\beta H/2)} 
\oplus \iota \frac{\exp (\beta H/2)}{2\cosh (\beta H/2)}
\iota^{-1}.  
\end{eqnarray*}
One sees that $\Gamma T_\beta
\Gamma + T_\beta = 1$ and $0\le T_\beta \le 1$, so it defines a 
quasifree state $\omega_\beta$ and by the GNS-construction a
representation $(\pi_\beta, \Omega_\beta, \bF_\beta)$ of $\bA$. 
The expressions for the modular Hamiltonian and the modular 
conjugation can be read of from 
\begin{eqnarray*}
\pi_\beta(X)^*\Omega_\beta = Je^{-\beta\,d\bF(\widetilde{H})/2} 
\pi_\beta(X) \Omega_\beta, 
\quad X \in \bA, 
\end{eqnarray*}
where
\begin{eqnarray*}
J(F_1 \wedge \dots \wedge F_n) \defeq \Gamma F_n \wedge \dots \wedge
\Gamma F_1, 
\end{eqnarray*}
and $d\bF$ is the second quantisation functor. This confirms that we 
have indeed defined a thermal equilibrium state at inverse 
temperature $\beta$, since the modular automorphism coincides with the
time-evolution of the system.  
\begin{thm}
The state induced by the operator $T_\beta$ is Hadamard in the sense
of Def. \ref{Hadamard}.
\end{thm}
\begin{proof}
From the definition of the spation temporal two-point functions,
Eq.~\eqref{twopointfunctions} and $T_\beta$ one calculates
\begin{eqnarray}\label{master}
\Lambda^{(\pm)}_\beta(v, u) = \langle \rho S\bar v, Q_\pm \rho Su 
\rangle, 
\end{eqnarray}
where
\begin{eqnarray*}
Q_\pm = \frac{\exp (\pm\beta H/2)}{2\cosh (\beta H/2)}.  
\end{eqnarray*}
Since $S$ is a solution 
to the Dirac equation in both entries, it is immediately clear that
\begin{eqnarray}\label{debbie}
Q_\pm\rho S = \rho q_\pm(i\partial_t)S 
\end{eqnarray}
modulo $\bL^{-\infty}$, 
where $q_\pm \in \bS^0(\bR)$ are given by
\begin{eqnarray*}
q_\pm(\lambda) = \frac{e^{\pm\beta\lambda/2}}{2\cosh(\beta\lambda/2)}.
\end{eqnarray*} 
Let $\chi_\pm$ be smooth functions on the real line
equal to the characteristic functions of $\bR_\pm$ except for a 
small neighbourhood of the origin. 
Now $\chi_\mp q_\pm$ is a symbol in $\bS^{-\infty}(\bR)$, hence
$\chi_\mp(i\partial_t)q_\pm(i\partial_t) \in \bL^{-\infty}(\bR)$ and   
we can conclude that (identifying the operators with the corresponding
distribution kernels)
\begin{eqnarray} \label{chiq}
\WF'(\chi_\mp q_\pm(i\partial_t)) \subset
\{(\vec x, t, \vec \xi, 0; \vec x, t, \vec \xi, 0): \quad 
(\vec x, \vec \xi) \in T^*\Sigma, \,\, t \in \bR \}.
\end{eqnarray}
By Thm.~\ref{wfprod} and Eq.~\eqref{chiq}, 
\begin{eqnarray*}
\WF'(\chi_\mp(i\partial_t)q_\pm(i\partial_t)S) &\subset&
\WF'(\chi_\mp q_\pm (i\partial_t)) \circ
\WF'(S) \\
&\cup& \WF_M(\chi_\mp q_\pm(i\partial_t)) \times M \\
&\cup& M \times \WF_M(S) = \emptyset, 
\end{eqnarray*}
implying $\chi_\mp(i\partial_t) q_\pm(i\partial_t) S \in
C^\infty$. On the other side, the leading symbol of 
$\chi_\pm(i\partial_t)$ is equal
to the characteristic function of the positive/negative real line.
Hence by definition of the Wave Front Set and the pseudo-local property
Eq.~\eqref{pseudolocal}
\begin{eqnarray}\label{wfqs}
\WF'(q_\pm(\partial_t)S) \subset \WF'(S) \cap (\chi_\mp^{-1}(0) \times M)
= C \cap (N_\pm \times N_\pm),
\end{eqnarray}
because $\WF'(S) = C$, which can be inferred from the corresponding
property of the scalar causal propagator \cite{Rad1}. 
By Thm.~\ref{rest} and Eqs.~\eqref{debbie} and 
\eqref{wfqs},  
\begin{eqnarray}\label{hsiao}
\WF'(Q_\pm \rho S) \subset (d\phi_1)^t(C \cap (N_\pm \times N_\pm)), 
\quad \WF'(S\rho') \subset (d\phi_2)^t(C), 
\end{eqnarray}
where $\phi_1: \Sigma \times M \rightarrow M \times M$ and 
$\phi_2: M \times \Sigma \rightarrow M \times M$ are the embeddings.
Taking into account Eq. \eqref{master}, we see that by Thm. \ref{wfprod}
\begin{eqnarray*}
\WF'(\Lambda^{(\pm)}_\beta) &=& \WF'(S\rho'Q_\pm\rho S) \subset
\WF'(S\rho') \circ \WF'(Q_\pm\rho S) \\
&\subset& (d\phi_2)^t(C)\circ (d\phi_1)^t(C \cap (N_\pm \times N_\pm)) = 
C \cap (N_\pm \times N_\pm). 
\end{eqnarray*}
To prove that equality holds in the above inclusion and that the
polarisation has the prescribed form, we can proceed as in proof of 
Thm. \ref{mainthm}. Hence we conclude that the state defined by $T_\beta$
is indeed Hadamard, i.e. the positive and negative two-point functions have
a Polarisation Set of the form Eq.~\eqref{77}. 
\end{proof}

\section{Adiabatic vacuum states}
   
It is known that ground states on general globally hyperbolic
space-times cannot be constructed in the same way as in static ones,
i.e. by just considering the spectrum of the Hamiltonian at an instant 
of time as above. This is related to the fact that there is no reasonable 
notion of positive and negative frequency modes in general globally 
hyperbolic space-times. Instead, they have to be 
extrapolated off the Cauchy surface. Following an idea by W. Junker
\cite{Junk}, this might be achieved by considering a factorisation of 
the (spinorial) Klein-Gordon operator. The states obtained are 
in a sense an approximation of a ground state at
some time $t$, taking the time-evolution of the metric in an 
infinitesimal neighbourhood of the Cauchy surface into account. 
They are called `adiabatic'. The failure of the `naive' ground state on
a globally hyperbolic space-time to describe a physical state  
may be seen more formally from the fact that it can be shown not to 
satisfy the Hadamard condition. 
We will now construct such an adiabatic ground state for the Dirac 
field. 

On each Cauchy surface $\Sigma(t)$, one can find an elliptic 
PDO $A(t)$ of order one satisfying
\begin{eqnarray}\label{foldy}
(-i n^\mu \nabla_\mu + A(t)) \circ 
(i n^\mu\nabla_\mu + A(t)) = g^{\mu \nu} \nabla_\mu 
\nabla_\nu + \frac{1}{4} R + m^2 \defeq P
\end{eqnarray}
modulo $\bL^{-\infty}$. A principal symbol of $A(t)$ is  
\begin{eqnarray}\label{symb}
a_1(\vec x, t, \vec \xi) = \sqrt{h(\vec \xi, \vec \xi)}.  
\end{eqnarray}
Such an operator may be constructed as an asymptotic expansion of its
symbol \cite{Junk}. We let $A^{-1}$ be a parametrix (we drop the 
reference to $t$ in places to lighten the notation). Then the
operator 
\begin{eqnarray*}
B \defeq \bRe (A^{-1} H)
\end{eqnarray*}
is a selfadjoint elliptic PDO of order zero with principal symbol
\begin{eqnarray*}
b_0(\vec x, t, \vec \xi) = a_1(\vec x, t, \vec \xi)^{-1}
h_1(\vec x, t, \vec \xi) = \frac{\mslash n \mslash \vec \xi}{
\sqrt{h(\vec \xi, \vec \xi)}}.
\end{eqnarray*}
By elementary continuity
properties of PDO's, it extends to a bounded, 
selfadjoint operator on $L^2(\Sigma, DM)$ and we can employ the spectral
calculus to define functions of this operator. Let $\chi_\pm$ be the 
characteristic functions of the positive resp. negative axis and
\begin{eqnarray*}
T_a \defeq \chi^+(B) \oplus \iota \chi^-(B)
\iota^{-1}. 
\end{eqnarray*}
Obviously, $T_a$ is a projection and $\Gamma T_a \Gamma = 1 - T_a$. 
It defines a pure quasifree and gauge invariant 
state $\omega_a$ on $\bA$. If $(M, g)$ is an ultrastatic 
space-time, then $A = |H|$. In that case, $T_a$ is the spectral projection 
corresponding to the positive
part of the spectrum of $H$, i.e. our state is the ground state.
In that sense our state should be understood as approximating the
ground state. We now state the main theorem of this paper.

\begin{thm}\label{mainthm}
The state defined by $T_a$ is Hadamard. 
\end{thm}
\begin{proof}
The expressions for the positive/negative frequent two-point 
functions are
\begin{eqnarray*}
\Lambda^{(\pm)}_a(v, u) = \left\langle \rho S \bar v, \chi^\pm(\bRe(A^{-1} H
)) \,\rho Su \right\rangle.
\end{eqnarray*}
In the first part of the proof we will show that the 
Wave Front Set of the above two-point functions has the required form, 
whereas in the second part we will verify the claims about the 
polarisation specified in Def. \ref{Hadamard}. 

Let us set $f^\pm(\lambda) = (1 \pm \lambda)^{-2} \chi^\pm(\lambda)$.
We first want to show that $f^\pm(B)$ are PDO's. To this end notice that
\begin{eqnarray*}
f^\pm(B)u = \frac{1}{2\pi i} \oint_{{\cal C}_\pm} \frac{1}{(1 \pm z)^2}
R(z, B)u\,dz, 
\end{eqnarray*}
where ${\cal C}_\pm$ is a contour around the positive part of the 
spectrum, keeping away from the set $[\varepsilon, C]$ resp. 
$[-C, -\varepsilon]$ for some $C, \varepsilon >0$, to be 
specified in a second. $R(z, B)$ is the 
resolvent of $B$. We can always choose such contours, because $B$ is
bounded and zero cannot be an accumulation point in the spectrum, the
latter fact following easily from the existence of a parametrix to
$B$. By Lemma \ref{technical_lemma} the resolvent is given 
by $r(z, \vec x, \vec D)$, where $r$ is a symbol in $\bS^0$ depending smoothly 
on $z$ in any open region of the complex plane whose closure does not
intersect $spec\,B \cup [-C, -\varepsilon] \cup [\varepsilon, C]$, for 
some $C, \varepsilon > 0$. 
Because $(1 \pm z)^{-2}$ has no pole for 
$\bRe z \ge 0$ resp. $\le 0$, it follows that $f^\pm(B)$ is the PDO corresponding
to the symbol
\begin{eqnarray*}
\sigma(f^\pm(B)) = 
\frac{1}{2\pi i} \oint_{{\cal C}_\pm} \frac{1}{(1 \pm z)^2}
r(z, \vec x, \vec \xi) \,dz, 
\end{eqnarray*}
Let us set
\begin{eqnarray*}
K_1^\pm = i S \rho' \bRe(1 \pm A^{-1} H)\rho
S, \quad K_2^\pm = \mslash n f^\pm(\bRe (A^{-1} H))\mslash n.
\end{eqnarray*}
Using the identity $\rho S \rho' = -i\mslash n$ (see Dimock,
\cite{Dim1}), it follows that we can write   
\begin{eqnarray*}
\Lambda^{(\pm)}_a(v, u) = \left\langle \rho K_1^\pm \bar v, 
K_2^\pm \rho K_1^\pm u \right\rangle. 
\end{eqnarray*} 
Suppose for the moment that the following inclusions hold:
\begin{eqnarray}
\label{hyp1}
&&\WF'(\Lambda^{(\pm)}_a) \subset \WF'(K_1^\pm \rho') \circ 
\WF'(K_2^\pm \rho K_1^\pm),\\
\label{hyp2}
&&\WF'(K_1^\pm) \subset 
C \cap (N_\pm \times N_\pm). 
\end{eqnarray}
As above, let $\phi_1: \Sigma \times M \to M \times M$ and $\phi_2: M \times 
\Sigma \to M \times M$ be the embeddings. It follows from the
pseudo-local property Eq.~\eqref{pseudolocal} of the operator $K_2^\pm$ 
and the behaviour of Wave Front Sets under restriction (Thm.~\ref{rest}) that  
\begin{eqnarray}\label{incl}
\WF'(\Lambda^{(\pm)}_a) &\subset& 
\WF'(K_1^\pm \rho') \circ \WF'(\rho K_1^\pm)\nonumber\\
&\subset& 
(d\phi_1)^t (C \cap (N_\pm \times N_\pm)) \circ
(d\phi_2)^t (C \cap (N_\pm \times N_\pm))\\
&=& C \cap (N_\pm \times N_\pm)\nonumber. 
\end{eqnarray}
The first inclusion Eq.~\eqref{hyp1} follows essentially from 
Thm.~\ref{wfprod}, but there are some technicalities. For details we refer the
reader to \cite{Junk}, where a similar statement is demonstrated. 
To prove the second inclusion Eq.~\eqref{hyp2}, we notice that
\begin{eqnarray*}
K_1^\pm = S \rho' A^{-1} \rho (A \pm i n^\mu\nabla_\mu) S + 
          S (A^* \pm i n^\mu\nabla_\mu) \rho' A^{*-1} \rho S. 
\end{eqnarray*} 
It follows from the Lichnerowicz formula for the square of the 
Dirac operator that
\begin{eqnarray}\label{Lichn}
(g^{\mu\nu}\nabla_\mu\nabla_\nu + \frac{1}{4}R + m^2)S = 0, 
\end{eqnarray}
both for the spinor and cospinor propagator. Now by Eq. \eqref{foldy}, 
the definition of the Wave-Front Set and the equality $\WF'(S) = C$ it 
holds that
\begin{eqnarray*}
\WF'((A \pm in^\mu \nabla_\mu)S) &\subset& C \cap 
\sigma_1(A \mp in^\mu\nabla_\mu)^{-1}(0)\\
\WF'(S(A^* \pm in^\mu \nabla_\mu)) &\subset& C \cap 
\sigma_1(A^* \mp in^\mu\nabla_\mu)^{-1}(0).
\end{eqnarray*}        
The main point here is that by Eq.~\eqref{symb} and the structure of $C$  
the right hand sides of the above inclusions are in fact equal
to $C \cap (N_\pm \times N_\pm)$. Also, 
\begin{eqnarray*}
\WF'(S\rho'A^{-1}) &=& \WF'(S\rho') \subset (d\phi_2)^t(C), \\ 
\WF'(A^{*-1}\rho S) &=&  \WF'(\rho S) \subset (d\phi_1)^t(C), 
\end{eqnarray*}
since $A$ is elliptic and by the restriction property of the
Wave Front Set. Using the above relations it follows
\begin{eqnarray*}
\WF'(S\rho'A^{-1}\rho(A \pm in^\mu\nabla_\mu)S) &\subset&
\WF'(S\rho'A^{-1}) \circ \WF'(\rho(A \pm in^\mu\nabla_\mu)S)\\
&\subset& (d\phi_2)^t(C) \circ (d\phi_1)^t(C \cap (N_\pm \times
N_\pm))\\
&=& C \cap (N_\pm \times N_\pm)
\end{eqnarray*}
and similarly for $\WF'(S(A^* \pm in^\mu\nabla_\mu)\rho'A^{*-1}\rho S)$.
This proves the inclusion Eq. \eqref{hyp2}. Now from the 
$\car$, Eq. \eqref{car} we infer that $\Lambda^{(+)}_a + \Lambda^{(-)}_a = iS$, hence
$C \subset \WF'(\Lambda^{(+)}_a) \cup \WF'(\Lambda^{(-)}_a)$. From this it follows
that in fact equality must hold in Eq. \eqref{incl}, i.e.
\begin{eqnarray*}
\WF'(\Lambda^{(\pm)}_a) = C \cap (N_\pm \times N_\pm).
\end{eqnarray*}
Looking at the
definition of Hadamard states, Def. \ref{Hadamard}, we see that only 
the polarisation remains to be verified. 

We aim at using the propagation of singularities theorem, 
Thm. \ref{propsing}. 
Since $\Lambda^{(\pm)}_a$ satisfy the Dirac equation, by the Lichnerowicz
formula Eq.~\eqref{Lichn}, they also satisfy 
\begin{eqnarray*} 
(P \otimes 1)\Lambda^{(\pm)}_a = 
(1 \otimes P)\Lambda^{(\pm)}_a = 0.
\end{eqnarray*}
$P$ is clearly of real principal type with $p_0(x, \xi) = -g(\xi, \xi)$.
We might set $\tilde p_0(x, \xi) = 1$, so the Dencker connection Eq. 
\eqref{dencon} corresponding to the operator $P$ becomes 
\begin{eqnarray*}
\bD_P = -2\xi^\mu\frac{\partial}{\partial x^\mu} + 
2\Gamma_{\mu\nu\sigma}\xi^\nu\xi^\sigma
\frac{\partial}{\partial \xi_\mu} 
+ \frac{1}{2}\gamma^a\gamma^b
\Gamma_{ab\mu}\xi^\mu - \Gamma_{\nu\mu}^\mu \xi^\nu. 
\end{eqnarray*}
To see the geometric meaning of the Dencker connection, we let
$\pi: T^*M \to M$ be the projection from the cotangent bundle to
the base. We may then rewrite the Dencker connection as the pull-back to 
$T^*M$ of the connection $\nabla + d\log|g|^{-1/4}$, the natural
connection in the bundle $DM \otimes L^{1/2}$, 
taken in the direction of the Hamilton vectorfield, 
\begin{eqnarray*}
\bD_P = \pi^*(\nabla + d\log|g|^{-1/4})_{\bH_q}. 
\end{eqnarray*}
Here $L^{1/2}$ is the line bundle of half-densities over $M$ and 
\begin{eqnarray*}
\bH_q = -2\xi^\mu\frac{\partial}{\partial x^\mu} + 
2\Gamma_{\mu\nu\sigma}\xi^\nu\xi^\sigma
\frac{\partial}{\partial \xi_\mu} 
\end{eqnarray*}
is the Hamilton vectorfield corresponding to $q(x, \xi) =-g(\xi, \xi)$, 
generating null-geodesics in $M$. Sections over integral curves of
$\bH_q$, annihilated by $\bD_P$ are thus pull-backs to $T^*M$ of
sections in $DM$ over null-geodesics which are parallel with respect to
$\nabla$. Hence, two elements $(x_1, \xi_1, u_1)$ and $(x_2, \xi_2, u_2)$ of
$\pi^* DM$ are in the same Hamiltonian orbit if 
$(x_1, \xi_1) \sim (x_2, \xi_2)$ and if $u_1 = \lambda
\bI_\nabla(x_1, x_2)u_2$ where $\bI_\nabla$ denotes parallel transport in 
the spin-bundle along a null-geodesic and $\lambda \in \bR$. 
From this one immediately finds the Hamiltonian
orbits corresponding to the operators $P \otimes 1$ and 
$1 \otimes P$. Let 
\begin{eqnarray*}
(x_1, \xi_1, x_2, \xi_2, w) \in T^*_{x_1}M \times 
T^*_{x_2}M \times D^*_{x_1}M \otimes D_{x_2}M
\end{eqnarray*}
be in the Polarisation
Set $\Pol'(\Lambda^{(\pm)}_a)$. Then if $w \neq 0$, the pair 
$(x_1, \xi_1, x_2, \xi_2)$ must be in $\WF'(\Lambda^{(\pm)}_a)$ which was 
shown to be equals $C\cap(N_\pm \times N_\pm)$. By the propagation of 
singularities theorem, Thm. \ref{propsing}, the 
Polarisation Set must be a union of Hamiltonian orbits. This implies that 
\begin{align}
(x_1, \xi_1, x_2, \xi_2, w) &\in 
\Pol'(\Lambda^{(\pm)}_a)\notag\\
\Leftrightarrow (x_1, \xi_1, x_1, \xi_1, w\bI_\nabla(x_2, x_1)) &\in 
\Pol'(\Lambda^{(\pm)}_a)\notag\\
\Leftrightarrow(x_2, \xi_2, x_2, \xi_2, \bI_\nabla(x_2, x_1)w) &\in 
\Pol'(\Lambda^{(\pm)}_a).\notag
\end{align}
Now since the 
Polarisation Set is an invariant object, $w\bI_\nabla(x_2, x_1)$ must
transform like a gamma matrix under a change of gauge, hence one concludes that
it must be proportional to $\mslash \xi_1$ and
$\bI_\nabla(x_2, x_1)w$ is proportional to $\mslash \xi_2$. 
But this is just the condition on the polarisation 
in Def. \ref{Hadamard}. 
\end{proof}

\section{Appendix}

In what follows we shall need various results and definitions from the
theory of distributions and the theory of pseudo-differential operators
(PDO's). 
If not indicated otherwise, these may be found in standard textbooks, for example
see \cite{tay, ho}. PDO's generalise classical
differential operators in the sense that they give meaning to fractional
powers of derivatives. They are defined in terms of so-called symbols. 
We shall not give the most general definition of a symbol here, since
only a certain class of symbols is relevant to this work. 

\begin{defn}
Let $\bO$ be a subset of $\bR^n$ and $m$ be a real number. Then one 
defines the symbols of order $m$ to be the set $\bS^m(\bO, \bR^n)$ of
all functions $a \in C^\infty(\bO, \bR^n)$ such that for every compact
subset $K$ of $\bO$ the following estimate holds 
\begin{eqnarray}\label{symbest}
\left|D^\alpha_x D^\beta_\xi a(x, \xi) \right|\le C_{\alpha, \beta, K}
(1 + |\xi|)^{m - |\beta|}
\end{eqnarray}
for all multi-indices $\alpha, \beta$. $D^\alpha$ is $i^{|\alpha|}
\partial_1^{\alpha_1} \dots \partial_n^{\alpha_n}$.  
One also writes $\bS^{-\infty} = \bigcap_m \bS^m$. 
\end{defn}
There is the notion of the asymptotic expansion of a 
symbol which is an important tool for constructing PDO's. 
Suppose $a_j \in \bS^{m_j}(\bO, \bR^n)$ for $j = 1, 2, \dots$ 
with $m_j$ monotonically decreasing to minus infinity. Then there exists
$a \in \bS^{m_0}(\bO, \bR^n)$ such that for all $N$ 
\begin{eqnarray*}
a - \sum_{j = 0}^N a_j \in \bS^{m_N}(\bO, \bR^n)
\end{eqnarray*}
and $a$ is defined modulo $\bS^{-\infty}$. 
If $a \in \bS^{m}(\bO, \bR^n)$ then the operator
\begin{eqnarray*}
Au(x) = \int e^{ix\xi} a(x, \xi) \hat u(\xi) \frac{d^n\xi}{(2\pi)^n}
\end{eqnarray*} 
is said to belong to $\bL^m(\bO)$, the PDO's of 
order $m$. For the operator $A$ one also writes $a(x, D)$. 
$A$ is a continous linear operator from $\bD(\bO)$ to $C^\infty(\bR^n)$. 
By the Schwartz kernel theorem it is thus given
by a distribution kernel $K_A \in \bD'(\bO \times \bO)$. $K_A$ is 
smooth off the diagonal in $\bO \times \bO$ and smooth everywhere 
in $\bO \times \bO$ if $A \in \bL^{-\infty}(\bO)$. Hence the asymptotic 
expansion of a symbol uniquely determines a PDO 
modulo smoothing operators. The above statement carries over
to matrix valued symbols and PDO's without 
major changes. If $a, b$ are symbols (possibly matrix valued), then
the convolution product is defined to be
\begin{eqnarray}
a \circ b(x, \xi) \sim 
\sum_{\alpha \ge 0} \partial^\alpha_x a(x, \xi) D^\alpha_\xi b(x, \xi)
/\alpha!
\end{eqnarray}
A PDO $A$ is said to be properly supported
if the support of its kernel distribution has compact intersection 
with any set of the form $\bO \times K$, $K \times \bO$, $K$ compact. 
If $A, B$ are properly supported PDO's, then 
$AB$ is a PDO with symbol $\sigma(AB) \sim 
\sigma(A) \circ \sigma(B)$.
The principal symbol $\sigma_m(A)$ of a pseudo-differential operator of order 
$m$ is the representer of its symbol in $\bS^m(\bO, \bR^n)/
\bS^{m-1}(\bO, \bR^n)$. 
It transforms under a change of coordinates in such a way as to
give a well-defined function on the cotangent bundle. By the composition
law for symbols, it behaves multiplicatively under multiplication of
two PDO's. If we have a smooth manifold $M$ instead of $\bR^n$ 
(or more generally a vector bundle $\bf E$), PDO's are defined to be the 
continuous operators on $\bD(M, {\bf E})$ which have the above properties in each
coordinte patch. A PDO $A$ on a vector-bundle $\bf E$ is called 
elliptic of order $m$, if its symbol $a$ is an invertible matrix for 
for $|\xi|$ greater than some constant and the estimate 
\begin{eqnarray*}
\|a(x, \xi)^{-1}\| \le C |\xi|^{-m}
\end{eqnarray*}
holds. Note that ellipticity follows from the corresponding property of 
any principal symbol alone. Elliptic PDO's are invertible in the
following sense:
\begin{thm}
If $A \in \bL^m$ is elliptic, then there exists a parametrix $B
\in \bL^{-m}$ for $A$ in the sense that $AB = BA = 1 \,\,
mod \,\, \bL^{-\infty}$. Such a parametrix is uniquely defined up
to $\bL^{-\infty}$.
\end{thm} 

We come to the definition of the Polarization Set of a vector-valued 
distribution $u = (u^1, \cdots, u^k) \in \bD'(\bO)^k$, $\bO$ an
open subset of $\bR^n$. For details of the definition and the subsequent
results see the paper by N. Dencker, \cite{Denck}.  
\begin{defn}
The Polarisation Set of a vector-valued distribution $u$ is defined as
\begin{eqnarray*}
\Pol(u) = \bigcap_{A \in \bL^0, \,\, Au \in C^\infty} \bN_A,  
\end{eqnarray*}
where $\bN_A$ is the set of all $(x, \xi, w) \in T^*\bO \times \bC^k$
such that $\sigma_0(A)(x,\xi)w = 0$. 
\end{defn}

PDO's are pseudo-local in the sense that 
they do not enlarge the Polarisation Set of a distribution, 
\begin{eqnarray}
\label{pseudolocal}
\Pol(Au) \subset \Pol(u).
\end{eqnarray} 

From the transformation properties of the principal symbol it is clear
that the definition can be carried over to the case of distributions
with values in a vector-bundle $\bf E$. $\Pol(u)$ is then seen to be a
linear subset of $\pi^* \bf E$, $\pi: T^*M \rightarrow M$ being the canonical 
projection in the fibres of the cotangent bundle. The ordinary Wave
Front Set $\WF(u)$ of a distribution is obtained by taking all points
$(x, \xi) \in T^*M$ such that the fibre over this point in
$\Pol(u)$ is nontrivial. We quote two results on the behaviour Wave-Front 
Set under composition and restriction important to this work:

\begin{thm}\label{rest} (Theorem 8.2.4 of \cite{ho}):
Let $u \in \bD'(M)$ and let $\phi: \Sigma \rightarrow M$ be a regularly 
embedded hypersurface in $M$. Then $u$ can be restricted to 
$\Sigma$ if $\WF(u) \cap N\Sigma = \emptyset$, $N\Sigma \subset T^*M$ 
being the conormal bundle to $\Sigma$. In this case 
$\WF(\rho u) \subset (d\phi)^t (\WF(u))$, where $\rho$ 
denotes restriction and $(d\phi)^t$ is the transpose of the 
tangent map.
\end{thm} 
\begin{thm}\label{wfprod} (Theorem 8.2.14 of \cite{ho}). 
Let $A: \bD(M_1) \rightarrow \bD'(M_2)$ and 
$B: \bD(M_2) \rightarrow \bD'(M_3)$ linear continous maps. 
By the Schwartz Kernel theorem these correspond to distribution
kernels $K_A \in \bD'(M_2 \times M_1)$ and $K_B \in \bD'(M_3 \times M_2)$.
If 
\begin{eqnarray*}
\WF'(K_B)_{M_2} \cap \WF'(K_A)_{M_2} = \emptyset
\end{eqnarray*}
then the convolution $B \circ A$ is well defined and
\begin{eqnarray*}
\WF'(K_{B\circ A}) &\subset& (WF'(K_B) \circ \WF'(K_A))\\ 
&\cup& (M_1 \times \WF(K_A)_{M_3}) \cup (\WF(K_B)_{M_1} \times M_3)
\end{eqnarray*} 
Here the prime means that one has to reverse the sign of
the corresponding cotangent vector in the second slot, 
and for $K \in \bD'(M_1 \times M_2)$,
\begin{eqnarray*}
\WF(K)_{M_2} = \{ (x_2, \xi_2): \quad (x_1, 0; x_2, \xi_2)
\in \WF(K)\}.
\end{eqnarray*}
\end{thm}

There is an important theorem on the Polarisation Set of distributions 
satisfying $Au \in C^\infty$ for differential operators $A$ of real 
principal type, which goes under the name `propagation of singularities'
\cite{Denck}. Such operators are defined as follows:
\begin{defn}
A $k \times k$ system $P$ of differential operators on 
a manifold $M$ with principal symbol $p_0(x, \xi)$ is said to be
of real principal type at $(y, \eta)$ if there exists a 
$k \times k$ symbol $\tilde p_0(x, \xi)$ such that 
\begin{eqnarray*}
\tilde p_0(x, \xi) p_0(x, \xi) = q(x, \xi)1_k
\end{eqnarray*} 
in a neighbourhood of $(y, \eta)$, where $q(x, \xi)$ is scalar and
real. 
\end{defn}
One sets
\begin{eqnarray*}
\bQ_P = \{ (x, \xi) \in T^*M - (0): \quad
det(p_0(x, \xi)) = 0 \}. 
\end{eqnarray*}
If $f$ is a $C^{\infty}$ function on $\bQ_P$ with values in
$\bC^k$, then one defines
\begin{eqnarray} \label{dencon}
\bD_Pf = \bH_qf + \frac{1}{2}\{\tilde p_0, p_0\} f + i\tilde p_0 p^s f, 
\end{eqnarray}
$\bH_q$ being the Hamiltonian vectorfield of $q$, 
\begin{eqnarray*}
\bH_q &=& \partial_x q \, \partial_\xi -
        \partial_\xi q \, \partial_x, \quad
\{ \tilde p_0, p_0 \} = \partial_\xi \tilde p_0 \, \partial_x p_0 -
                        \partial_x \tilde p_0 \, \partial_\xi p_0,\\
p^s &=& p_1 + 
\frac{1}{2}\partial_\xi D_x p_0, \quad
\sigma(P) \sim p_0 + p_1 + p_2 + \dots.
\end{eqnarray*} 
One can prove that $\bD_P$ is a partial connection along the 
Hamiltonian vectorfield  restricted to $\bQ_P$. 
Since there is some arbitrariness in the choice of the symbol $\tilde p$,
the partial connection is not uniquely defined. One can however 
prove that the remaining arbitrariness is irrelevant in what follows. 
\begin{defn}
A Hamilton orbit of a system $P$ of real principal type is 
a line bundle $\bL_P \subset \bN_P \vert c$, where $c$ is an
integral curve of the Hamiltonian field on $\bQ_P$ and 
$\bL_P$ is spanned by the sections $f$ satisfying $\bD_Pf = 0$, 
i.e. $\bL_P$ is parallel with respect to the partial connection. 
\end{defn}
The following theorem goes under the name 'propagation of singularities'
\cite{Denck, hodu}. 
\begin{thm}\label{propsing}
Let $P$ be as above and $u$ a vector-valued 
distribution. Suppose $(x, \xi) \notin \WF(Pu)$. Then, over a 
neighbourhood of $(x, \xi)$ in $\bQ_P$, $\WF_{pol}(u)$ is a
union of Hamilton orbits of $P$. 
\end{thm}

In this paper, we use the resolvent $R(z, Q) = (Q - z)^{-1}$ of a 
an elliptic selfadjoint PDO of order zero (always assuming that 
$z$ is not in $spec\,Q$). Let $U_{\varepsilon, C}$ be an open region in 
$\bC$ whose closure does not intersect the set 
$[-C, -\varepsilon] \cup [\varepsilon,
C] \cup spec\,Q$. The following lemma is needed in this work.  

\begin{lem}\label{technical_lemma}
Let $Q = q(x, D) \in \bL^0(M, \bE)$ be elliptic and self-adjoint, where $\bE$ is a 
vector bundle over $M$. Then $R(z, Q) = r(z, x, D)$ for some symbol 
$r \in \bS^0$, depending smoothly 
$z \in U_{\varepsilon, C}$ for some $\varepsilon, C > 0$.  
\end{lem} 
\begin{proof}
For simplicity, we shall only treat the case when ${\bf E} = \bO
\times \bC^k$ and $\bO$ an open subset in $\bR^n$. 
The proof can be adapted to the general case by using 
a partition of unity on $M$ and working in local trivialisations.
Suppose in the following that $z \in U_{\varepsilon, C}$ 
is as described above. By choosing a suitable principal symbol, we can 
assume that $\|q_0(x, \xi)^{-1}\|$ is bounded for all $\xi$ because $Q$ is
elliptic. Then $(q_0 - z)^{-1}$ will be in 
$\bS^0(\bO \times U_{\varepsilon, C},\bR^n)$ (we treat $z$ like
the variables $x$) for some $\varepsilon, C > 0$. If
\begin{eqnarray*}
(q - z)(x, D)(q_0 - z)^{-1}(x, D) = 1 - a_z(x, D), 
\end{eqnarray*}
then
\begin{eqnarray*}
\|\partial^l_z\partial_x^\alpha \partial_\xi^\beta a_z(x, \xi)\| 
\le C_{l, \alpha, \beta}
(1 + |\xi|)^{-1-|\beta|}
\end{eqnarray*}
Let $e_z(x, \xi)$ be an asymtpotic sum of the symbols of 
\begin{eqnarray*}
(q_0 - z)^{-1}(x, D)(a_z(x, D))^N, \quad N = 0,1,\dots.
\end{eqnarray*}
depending thus smoothly on $z\in U_{\varepsilon, C}$. Then one has 
\begin{eqnarray*}
(q - z)(x, D)e_z(x, D) = 1 - w_z(x, D), 
\end{eqnarray*}
and 
\begin{eqnarray}\label{westimate}
\|\partial_z^l\partial_x^\alpha \partial_\xi^\beta w_z(x, \xi)\| 
\le C_{l, \alpha, \beta}
(1 + |\xi|)^{-N-|\beta|}
\end{eqnarray}
for any $N$. Multiplication with $R(z, Q)$ gives
\begin{eqnarray}\label{rdef}
R(z, Q) = e_z(x, D) + R(z, Q)w_z(x, D).
\end{eqnarray}
Now by the estimate Eq. \eqref{westimate}, 
$w_z$ is a continuous map from $H^s(\bO, \bC^k)$ to $H^t(\bO, \bC^k)$, 
uniformly in $z$ and for any $t, s$, where we mean the Sobolev spaces. 
On the other hand, $R(z, Q)$ is a continuous operator in $H^t(\bO, \bC^k)$
uniformly in $z$, at least for $t = 0, 1, \dots$. 
This is obvious for $t = 0$. For $t > 0$ one has, choosing an
arbitrary positive elliptic operator $L$ of order $t$, 
\begin{eqnarray*}
\|R(z, Q) u\|_t &\le& C\|LR(z, Q) u\|_0 \le C
\|(Q - z)LR(z, Q)u\|_0 \\ 
&\le& C(\|u\|_t + \|[L,Q]R(z, Q)u\|_0) \le C(\|u\|_t + 
\|R(z, Q)u\|_{t-1}), 
\end{eqnarray*}
since $[L,Q]$ is a PDO of order $t-1$. The result follows by induction
on $t$. Hence, for any $t, s$, the operator $R(z, Q)w_z(x,D)$ is continuous 
from $H^s(\bO, \bC^k)$ to $H^t(\bO, \bC^k)$, uniformly in $z$. 
It follows from well-known embedding theorems that it must hence be an
operator defined by a kernel in $C^\infty(\bO \times \bO, \bC^k
\times \bC^k)$, i.e. an operator in $\bL^{-\infty}$. The same can 
be shown to hold for any $z$-derivative of this operator using
the above estimates. Hence $R(z, Q)$ is a PDO with symbol $r
\in \bS^0(\bO \times U_{\varepsilon, C}, \bR^n)$, defined by 
Eq.~\eqref{rdef}. In particular, the symbol depends smoothly on
$z \in U_{\varepsilon, C}$. 
\end{proof}

\end{document}